*Article*

# From Coordination to Personalization: A Trust-Aware Simulation Framework for Emergency Department Decision Support

Zoi Lygizou * and Dimitris Kalles *

School of Science & Technology, Hellenic Open University, 26 335 Patra, Greece
* Correspondence: zoi.lygizou@ac.eap.gr (Z.L.); kalles@eap.gr (D.K.)

**Abstract**

**Background/Objectives**: Efficient task allocation in hospital emergency departments (EDs) is critical for operational efficiency and patient care quality, yet the complexity of staff coordination poses significant challenges. This study proposes a simulation-based framework for modeling doctors and nurses as intelligent agents guided by computational trust mechanisms. The objective is to explore how trust-informed coordination can support decision making in ED management. **Methods**: The framework was implemented in Unity, a 3D graphics platform, where agents assess their competence before undertaking tasks and adaptively coordinate with colleagues. The simulation environment enables real-time observation of workflow dynamics, resource utilization, and patient outcomes. We examined three scenarios – Baseline, Replacement, and Training – reflecting alternative staff management strategies. **Results**: Trust-informed task allocation balanced patient safety and efficiency by adapting to nurse performance levels. In the Baseline scenario, prioritizing safety reduced errors but increased patient delays compared to a FIFO policy. The Replacement scenario improved throughput and reduced delays, though at additional staffing cost. The training scenario fostered long-term skill development among low-performing nurses, despite short-term delays and risks. These results highlight the trade-off between immediate efficiency gains and sustainable capacity building in ED staffing. **Conclusions**: The proposed framework demonstrates the potential of computational trust for evidence-based decision support in emergency medicine. By linking staff coordination with adaptive decision making, it provides hospital managers with a tool to evaluate alternative policies under controlled and repeatable conditions, while also laying a foundation for future AI-driven personalized decision support.

**Keywords:** Simulation-based decision support; Computational trust models; Task allocation; Emergency Department (ED); Healthcare staffing and coordination; Multi-agent systems; Workflow optimization; Patient safety; Resource utilization

## 1. Introduction

Emergency Departments (EDs) are among the most overcrowded units in modern healthcare systems, and overcrowding has been recognized as a global challenge for more than two decades [1,2]. Patients arrive with highly diverse medical conditions and varying levels of acuity, yet EDs are obligated to provide care at the highest possible standard for all. Errors in ED procedures can have severe consequences, including disability or death. At the same time, the majority of patients – particularly within public EDs – face long lengths of stay, while hospital managers struggle with persistent budgetary and staffing constraints [3–6].

Public hospitals often operate under multiple, and sometimes conflicting, constraints such as limited budgets, shortages of skilled personnel, and restricted access to medical equipment. These challenges make it difficult to ensure both patient safety and operational efficiency. Because EDs operate continuously, including holidays, it is practically impossible to interrupt or reconfigure real ED workflows for the sole purpose of evaluating performance and efficiency [7–9]. Recent studies have explored partial flexibility in staffing, where nurses or servers can be reassigned at discrete intervals, such as the beginning of shifts, to better match demand while respecting operational constraints [10,11]. Additionally, simulation and machine learning – based approaches have been shown to optimize resource scheduling, reduce patient length of stay, and improve overall operational efficiency in EDs [12,13]. As a result, hospital managers require robust decision-support tools that enable evidence-based evaluation of alternative policies and interventions. Such tools should enable managers to systematically explore trade-offs between patient safety, operational efficiency, and staff development under controlled, repeatable conditions – ultimately allowing them test potential solutions before implementing changes in real ED environments [3,4,14,15].

Beyond improving operational efficiency, such decision-support frameworks are directly relevant to the goals of personalized medicine [16]. In the emergency department setting, personalization extends not only to diagnosis and treatment, but also to the allocation of healthcare resources in real time. Recent work has shown how machine learning can be used to personalize patient flow pathways and improve throughput in the ED [10]. Each patient arrives with unique clinical needs and acuity levels, and delays in care – caused, amongst a variety of reasons, by suboptimal task allocation or staffing mismatches – translate into heterogeneous risks at the individual level. By integrating computational trust models with predictive AI techniques, it becomes possible to dynamically adapt the distribution of staff and tasks to match the evolving needs of specific patients and situations. In this way, system-level optimization contributes directly to patient-specific outcomes, aligning workflow efficiency with the principles of personalized medicine. Future extensions of our framework could further strengthen this alignment by incorporating patient-level data (e.g., triage information, medical histories, or real-time vital signs) into the trust-based allocation process, enabling both personalized staffing strategies and individualized clinical care.

In dynamic work environments such as a hospital emergency department, the problem of task allocation closely resembles challenges observed in open multi-agent systems (MASs), where agents must self-organize under uncertainty and resource constraints. Modern computational trust and reputation models have been successfully applied to such problems in diverse domains, including peer-to-peer networks, online marketplaces, and the Internet of Things. In our previous work, we introduced the biologically inspired *Create Assemblies* (CA) trust model, which draws inspiration from synaptic plasticity in the human brain [17]. Unlike traditional models where the trustor (service requester) selects a trustee (service provider), CA enables the trustee to autonomously assess whether it possesses the necessary skills to undertake a task. More specifically, task requests are broadcast to potential trustees, which locally store the request and update their trust weights based on performance feedback after task completion. This event-driven mechanism supports adaptive task allocation while preserving efficiency and robustness. Subsequently, we improved CA by incorporating self-assessment: after providing a service, a trustee re-evaluates its performance, and if it falls below a predefined threshold, it (auto-)classifies itself as an unreliable provider [18]. This modification ensures that each provider maintains an up-to-date evaluation of its own capabilities, enabling the immediate detection of performance drops that could be harmful.

Motivated by these foundations, in this work we proposed a simulation-based framework that leverages computational trust models to support decision making in emergency department staffing and workflow coordination. The framework, implemented in Unity, enables the modeling of doctors and nurses as intelligent agents operating under high variability and uncertainty, whose interactions are guided by trust-based task allocation mechanisms. Unlike traditional rule-based simulations, our approach focuses on capturing the dynamics of interpersonal trust and its impact on team performance, patient safety, and resource utilization.

By embedding trust models into agent decision-making, the system allows us to explore how different staffing configurations and training interventions affect both operational efficiency and patient-centered outcomes. In particular, the framework supports the evaluation of alternative policies under controlled and repeatable conditions, thereby offering hospital managers a tool for evidence-based experimentation without disrupting real-world clinical operations.

A key contribution of this work lies in linking staff coordination with the principles of personalized medicine. We demonstrate how adaptive task allocation, informed by computational trust, can evolve toward patient-specific care by dynamically aligning staff expertise and reliability with individual patient needs. This positions our simulation not only as a tool for operational research, but also as a stepping stone toward AI-driven, personalized decision-support in emergency medicine.

The remainder of this paper is organized as follows: Section 2 reviews related work across four key areas: simulation of EDs, decision-support systems in healthcare, computational trust models, and personalized medicine with adaptive staffing. Together, these perspectives provide the context and motivation for the proposed Unity-based simulation framework which is presented in Section 3. Section 4 introduces a case study with experiments comparing baseline, replacement, and training scenarios. Section 5 discusses results and limitations, while Section 6 concludes and outlines future research directions.

## 2. Related Work

Research on emergency department operations spans multiple methodological traditions, ranging from simulation-based approaches to decision-support systems and adaptive frameworks inspired by personalized medicine. Simulation studies have modeled patient flows, staffing policies, and resource constraints, while decision-support systems increasingly incorporate machine learning and optimization to enhance real-time management. At the same time, computational trust models –though widely applied in multi-agent systems – have only recently begun to appear in healthcare research, leaving intra-organizational collaboration and task allocation largely unexplored. Finally, emerging work on personalized medicine highlights the need to align operational decision-making with patient-specific characteristics. In this section, we review four key areas of related work: simulation of emergency departments, decision-support systems in healthcare, computational trust models, and personalized medicine and adaptive staffing. Together, these perspectives provide the foundation for our proposed framework, while also revealing critical gaps at the intersection of trust, adaptive decision-making, and patient-centered resource allocation.

### 2.1. Simulation of Emergency Departments

Simulation has become a well-established tool for studying Emergency Department (ED) operations, providing insights into patient flow, resource allocation, and operational bottlenecks. Among the most commonly used approaches are Discrete-Event Simulation (DES), Agent-Based Simulation (ABS), and System Dynamics (SD). Each method offers unique advantages and limitations in modeling complex healthcare systems.

DES has been extensively applied to model patient arrival processes, service times, and resource constraints. Studies using DES have demonstrated its effectiveness in analyzing patient flow, predicting waiting times, and evaluating interventions aimed at reducing overcrowding. For instance, Hamza et al. [3] proposed SIM-PFED, a hybrid DES-ABS model integrated with a multi-criteria decision-making approach (TOPSIS) to optimize patient flow and reduce throughput time in EDs, demonstrating significant reductions in waiting time and length of stay on real hospital data. Similarly, Kim [12] integrated DES with machine learning to dynamically allocate physicians, reducing patient length of stay, while Castanheira-Pinto et al. [13] combined DES with lean healthcare principles to identify bottlenecks and redesign a public hospital ED. Other works, such as De Santis et al. [5], have focused on DES model calibration to recover missing service parameters, and Dosi et al. [6] demonstrated how DES can be integrated with design thinking to guide organizational changes in ED operations. Collectively, these studies highlight DES as a robust and flexible tool for modeling ED processes, though it often abstracts away the nuances of individual staff behaviors and interpersonal interactions.

ABS extends traditional simulation by modeling each actor in the ED – patients, doctors, and nurses –as autonomous agents with individual attributes and decision-making rules. ABS allows for the exploration of how heterogeneous behaviors and interactions influence overall system performance. Beyond traditional ABS, recent work has explored how agent-based architectures can support Digital Twin (DT) frameworks for EDs. For example, Moyaux et al. [4] proposed an ABS-driven DT that represents patients, staff, and resources as synchronized agents, enabling real-time monitoring (Digital Shadow), predictive simulations of current states (Synchronized DT), and exploratory Monte Carlo "what-if" analyses (Exploratory DT). Their study highlights how ABS can evolve into a powerful tool not only for simulating operations but also for proactive, data-driven decision support in complex healthcare environments. Similarly, Mesas et al. [15] introduced a customizable

ABS framework that decomposes ED models into modular components, improving adaptability across different hospital contexts and operational constraints. Implemented in NetLogo and Python, their framework supports reusable, flexible simulation environments that facilitate "what-if" analysis and resource planning. Along the same line, Godfrey et al. [7] proposed a domain-specific modeling language (DSML) that enables reusable, stakeholder-friendly ABS models. Developed with National Health Service (NHS) professionals, their approach enhances credibility and engagement by allowing healthcare staff to directly contribute to model design, and demonstrates flexibility across multiple case studies. Recent work by Xuan et al. [8] further illustrates the breadth of ABS applications by examining how spatial design influences nursing efficiency across departments. Their study highlights that standardized layouts often neglect department-specific workflows, leading to inefficiencies and dissatisfaction, and shows how agent-based modeling can capture dynamic nursing behaviors under varying spatial configurations. Alongside this, Yuan and Zhou [9] integrate ABS with Space Syntax Analysis (SSA) to investigate how crowding impacts visibility in EDs, a factor crucial for surveillance, staff collaboration, and patient way finding. Their findings reveal that corridor layouts strongly influence congestion and natural surveillance, with ring configurations improving collaboration and simpler layouts benefiting navigation. Together, these studies demonstrate that ABS can extend beyond operational flow modeling to incorporate human-centered and contextual factors in healthcare delivery. This line of work emphasizes the growing versatility of ABS approaches, extending their applicability beyond rigid, monolithic models. However, despite these advances, existing ABS studies in EDs largely focus on task execution, operational efficiency, and structural adaptability, without deeply considering the dynamics of trust, collaboration, or adaptive decision-making among staff. Complementary work in healthcare simulation, such as Evans et al. [19], has highlighted the importance of team cognition – encompassing shared mental models and situational awareness – and proposed a deliberate framework for its measurement. Yet, such perspectives remain underrepresented in ED-focused simulation research, where operational models rarely capture the underlying cognitive and relational processes that drive team performance.

SD approaches provide a high-level perspective, capturing the feedback loops and accumulations inherent in ED operations. SD models have been used to study patient inflow-outflow dynamics, bottleneck identification, and policy evaluation at the organizational level. For instance, McAvoy et al. [20] developed an SD model of ED patient flow that incorporated ambulance and walk-in arrivals, acute and fast-track processes, diagnostic services, and short-stay units. Their model, validated with historical data, demonstrated which levers –such as reducing ambulance ramping – effectively alleviate congestion, while also revealing the limits of strategies like simply adding staff. Similarly, England et al. [21] developed an SD decision support tool to evaluate interventions for older ED patients, modeling the patient journey through admission, discharge, and potential readmission. Their simulations of five evidence-based interventions showed reductions in admissions, readmissions, and hospital-related mortality, highlighting the utility of SD for comparing care pathways and guiding clinical decision-making. Expanding the focus to intra-organizational dynamics, Wong et al. [2] developed a qualitative SD model examining the interplay of workplace violence, clinician burnout, and agitation management in EDs. Their model revealed feedback loops in which staff stress and burnout influence restraint decisions and patient outcomes, while mutual trust and team support mitigate negative cycles. In a resource-limited setting, Muttalib et al. [14] applied Group Model Building (GMB), a participatory SD approach, to identify barriers and facilitators of acute pediatric care delivery in a Malawian tertiary hospital. Their workshops highlighted severe illness and high patient volume as central stressors creating reinforcing feedback loops, while factors like parental engagement and provider resilience served as balancing elements. This study demonstrates the value of participatory SD methods in fostering shared understanding and identifying actionable interventions, even under constrained resources. Although SD can reveal structural issues affecting system performance, it generally lacks the granularity needed to capture individual-level variability or personalized interventions.

Overall, the literature demonstrates that simulation is a well-established and valuable tool for improving ED efficiency and patient flow. Yet, most existing work focuses on operational metrics and structural optimization, while interpersonal and team dynamics, as well as patient-specific decision-making, remain underexplored. Our work addresses this gap by incorporating computational trust models and agent-specific behaviors, providing a simulation framework that supports not only operational efficiency but also personalized, adaptive staffing strategies in line with emerging approaches in personalized medicine.

*2.2. Decision Support Systems in Healthcare*

Hospital managers increasingly rely on Decision Support Systems (DSS) to plan resources, schedule staff, and optimize patient flow. Approaches range from queuing theory and mathematical optimization to AI-driven scheduling algorithms.

For example, Yousefi and Yousefi [1] proposed a metamodel-based simulation optimization framework for staffing allocation in a Brazilian ED, combining an ensemble metamodel - integrating Adaptive Neuro-Fuzzy Inference System (ANFIS), Feed Forward Neural Network (FNN), and Recurrent Neural Network (RNN) with Adaptive Boosting (AdaBoost) – with a discrete Imperialist Competitive Algorithm (ICA), achieving a 24.82% reduction in door-to-doctor time. Similarly, Kim [12] integrated DES with machine learning to dynamically allocate physicians of varying experience levels in Korean EDs reducing patient length of stay while selecting optimal scheduling policies in real time. Similarly, Ortiz-Barrios et al. [22] combined AI with DES to address nurse staffing shortages in Spanish EDs during seasonal respiratory disease outbreaks. Their framework used Extreme Gradient Boosting (XGBoost) to predict treatment probabilities and integrated these forecasts into a simulation model, achieving high predictive accuracy and enabling data-driven staffing adjustments that reduced patient waiting times by up to 7.5 hours. This study illustrates the potential of coupling predictive analytics with simulation for managing demand surges in critical healthcare contexts. Complementing these algorithmic approaches, Castanheira-Pinto et al. [13] combined DES with lean healthcare principles to model, assess, and redesign a public hospital ED in Portugal, identifying bottlenecks, optimizing staffing requirements, and supporting evidence-based process improvements. De Santis et al. [5] addressed the challenge of incomplete timestamp data in DES models, proposing a simulation-based optimization approach that leverages derivative-free optimization and Weilbull distributions to accurately recover service times and critical patient cases, enhancing model reliability. Extending this line of research, Wang et al. [23] introduced a two-stage simulation-based optimization framework that incorporates part-time work shifts alongside traditional fixed schedules. Their approach uses nonlinear stochastic programming and an adaptive empirical stochastic branch-and-bound algorithm to determine optimal staffing levels, followed by integer programming for feasible shift assignments. Applied to ED settings, this framework demonstrated that integrating part-time scheduling during peak hours can significantly reduce patient waiting times, increase flexibility, and alleviate resource shortages without overburdening full-time staff. Similarly, Redondo et al. [24] focused on long-term capacity planning, proposing a simulation-optimization methodology that defines robust staffing levels under demand uncertainty. Their scenario-based Sample Average Approximation approach aggregates workload requirements across care pathways and resources, enabling managers to align future capacity with fluctuating patient needs over extended horizons. Beyond purely algorithmic interventions, Dosi et al. [6] demonstrated how combining design thinking with DES can guide participatory organizational changes in EDs, enabling measurable improvements in patient waiting times and staff work quality within 18 months.

Together, these studies illustrate the potential of simulation – and AI – based DSS to enhance operational efficiency, optimize staffing and scheduling decisions, and support real-time, data-driven management in complex ED environments. However, most existing DSS frameworks do not fully account for uncertainty in staff behavior, variations in skill levels, or interpersonal trust relationships, which can critically affect performance in real ED settings.

*2.3. Computational Trust Models*

Computational trust models, originate from computer science and multi-agent systems, with applications in e-commerce, collaborative robotics, and distributed decision-making. These models formalize trust relationships between agents to support more reliable coordination under uncertainty. Despite their widespread use in other domains, applications of computational trust in healthcare remain rare. Recent work by Anjana and Singhal [25] begins to address this gap by simulating trust dynamics in healthcare networks through a multi-agent, game-theoretic approach. Their study models interactions among hospitals, doctors, researchers, insurance companies, and patients in an Electronic Medical Record (EMR)-sharing environment, demonstrating how trust relationships evolve toward equilibrium through repeated cooperation and defection strategies. While this work highlights the relevance of computational trust for policy-making and system design in healthcare ecosystems, it primarily focuses on inter-organizational trust management. By contrast, intra-organizational dynamics such as staff collaboration and task allocation in EDs remain largely unexplored, leaning a significant gap in simulation-based decision-support research.

*2.4. Personalized Medicine and Adaptive Staffing*

Personalized medicine traditionally focuses on tailoring diagnosis and treatment to individual patient characteristics. Recent research, however, has begun to extend this concept to healthcare operations, exploring patient-specific workflow optimization and adaptive staffing strategies. Savchenko and Bunimovich-Mendrazitsky [26] investigate the economic feasibility of personalized medicine for healthcare service providers, using bladder cancer as a case study, and propose a framework for identifying patient cohorts to balance clinical effectiveness with operational efficiency. Building on this idea, Gummadi [16] proposes a multimodal AI framework that integrates Electronic Health Records, patient-reported outcomes, genomic data, and real-time physiological information from wearable sensors to create comprehensive patient profiles. This approach not only supports personalized treatment but also informs workflow optimization and staffing decisions, demonstrating improvements in clinical outcomes, patient satisfaction, and provider efficiency.

In the context of emergency care, Hodgson et al. [10] applied AI and machine learning to optimize ED patient flow through a personalized vertical processing pathway, showing how individualized risk scores can support flexible, patient-specific care protocols that enhance throughput without compromising safety. Complementing such AI-driven interventions, Chan et al. [11] study dynamic server assignment in multiclass queueing systems with shift-based reassignment constraints, motivated by nurse staffing in EDs. Their analysis demonstrates that partial flexibility – where reassignment is possible only at shift boundaries – can substantially reduce waiting costs compared to static staffing policies, while remaining operational feasible. Together, these studies demonstrate that personalized, patient-centered approaches can be extended beyond clinical interventions to influence operational planning, resource allocation, and adaptive staffing in complex healthcare environments.

Despite these advances, most existing frameworks focus on predictive or reactive decision-making, leaving a gap for simulation-based approaches that combine personalized patient data with computational trust and adaptive staffing models to optimize both efficiency and team performance in real-time. This study addresses these gaps by introducing a Unity-based simulation framework that incorporates computational trust models to explore their potential impact on ED efficiency, staff development, and patient-centered outcomes.

## 3. Materials and Methods

*3.1. Simulation Framework in Unity*

3.1.1. Overview

To systematically study and evaluate different decision-making strategies and behavioral patterns in a clinical context, we developed a simulation framework using the Unity game engine. The environment models the daily workflow of hospital staff, focusing particularly on the interactions between doctors, nurses, and patients within an emergency room (ER) setting. The Unity-based implementation enables the execution of realistic, real-time agent behaviors and facilitates the comparison of alternative policy scenarios in a controlled and repeatable manner.

3.1.2. Agents and Roles

The simulation comprises four primary agent categories:

- Doctors – evaluate patients and generate task requests (e.g., inserting an IV catheter), according to their assigned evaluation style. Each task request includes an estimated difficulty level and expected execution duration, derived from the doctor's assessment of the patient's condition;
- Nurses – receive and prioritize task requests based on a configurable decision policy (e.g., CA trust model, First-In-First-Out (FIFO)). Nurses then evaluate whether to accept or ignore each request. Their performance in executing accepted tasks is affected by their skill level (nurse quality) and in, certain scenarios, their training history. Task execution times are further influenced by the difficulty level assigned to the task and the nurse's capability;

- Trainer Nurses – specialized agents assigned to train low-performing nurses in Scenario 3. They accompany the trainee during task execution, enabling skill improvement through repeated observation and guidance;
- Patients, who are dynamically spawned and assigned to beds, each requiring a specific task to be completed.

The specific logic governing how doctors determine task difficulty (evaluation style) and how nurse skill level affects task execution time (nurse quality) is described in detail as follows:

**Doctor Evaluation Styles**

In the simulation, each doctor is assigned a predefined evaluation style at the start of the shift. The evaluation style determines how the doctor estimates the performance level required to perform a specific task after examining a patient.

Three styles are implemented:

- Estimates Correctly – the doctor accurately assesses the true performance level;
- Overestimates – the doctor systematically assigns higher performance levels than the true value;
- Underestimates – the doctor systematically assigns lower performance levels than the true value.

When a doctor examines a patient, the function EvaluatePerformanceLevel() maps the task's true difficulty to the estimated level according to the doctor's evaluation style. The pseudocode of the core decision logic is presented in Algorithm 1 (see Appendix A.1). This design enables the simulation to model different diagnostic behaviors, allowing us to study how misestimating of task difficulty affects nurse decision-making, workload distribution, and overall system performance.

Although these evaluation styles are predefined in the current implementation, in a real-world emergency room, such decision-making processes may vary considerably across individual doctors and over time. Approaches for learning personalized evaluation functions from empirical data, potentially using machine learning, are discussed in the Future Work Section.

**Nurse Quality Attributes**

At the start of each simulation run, nurses are instantiated with predefined quality attributes, currently set to either high-performing or low-performing. This attribute determines their expected task execution time and, consequently, their likelihood of task success. Nurse performance is modeled in the GetTaskDuration() function. Here, the nurse's quality influences the execution time for each task, with high-performing nurses generally completing tasks faster and with higher consistency, and low-performing nurses showing slower and more variable execution times. In training scenarios, low-performing nurses can improve their performance by performing tasks under the instructions of a trainer nurse, represented by an incremental bonus chance in their task duration calculation. The logic governing tasks execution speed for different nurse qualities is summarized in Algorithm 2 (see Appendix A.2.).

Such an approach allows us to systematically simulate the impact of different nurse skill levels on task allocation, patient service times, and overall ER performance. However, as with doctor evaluation styles, in a real-world emergency room, nurse performance is influenced by a complex interplay of factors, including experience, training, fatigue, workload, and individual skill variations. In the Future Work Section, we discuss the potential of using machine learning to learn this nurse performance model directly from empirical data, enabling the simulation to more accurately reflect real-world variability and adaptive behaviors.

3.1.3. ER Layout and Environment Setup

The emergency room (ER) is implemented as a structured 3D environment within Unity, designed to support realistic agent navigation and interactions.

Figure 1 illustrates the layout, which includes the following key areas:

- Bed Zone – A green colored floor area where patient beds are located and treatments are performed. The ER contains nine available beds. Each doctor has exclusive access to a fixed set of beds: Doctor with ID 1 is assigned to Beds 1-3, Doctor with ID 2 to Beds 4-6, and Doctor with ID 3 to Beds 7-9. This allocation constrains patient assignments and task requests to the relevant doctor-bed pairs;

- Staff Waiting Room – An orange-colored designated area where doctors and nurses are initially spawned at the start of the simulation and remain until they are assigned to patient care in the Bed Zone. Nurses also return to this area when no pending tasks are available;
- ER Exit Point – A tagged location (ERExit) where agents, such as patients who have completed treatment, navigate when leaving the ER.
- UI Overlays – Floating labels and task timers are attached to agents to display real-time information such as agent IDs and remaining task durations.

While the current implementation adopts a fixed number of beds, a predefined spatial arrangement, and static bed allocations to doctors, possible extensions for dynamically generated ER layouts and site-specific operational workflows are discussed in Future Work Section.

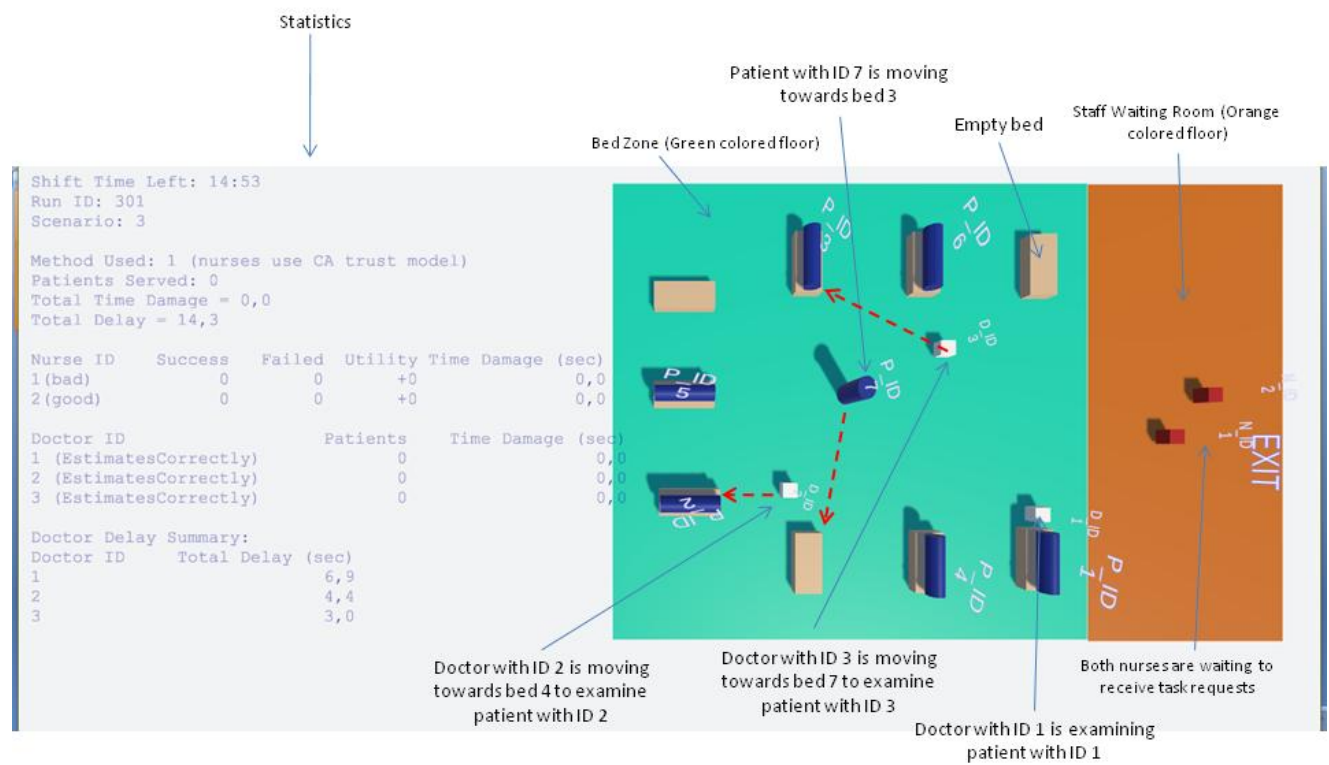

**Figure 1.** Simulation snapshot showing the bed zone (green area), staff waiting room (orange area), and the movement of agents. Red arrows indicate movement directions, while labels describe the current activity of agents. Dynamically updated statistics are displayed on the left side.

3.1.4. Scenarios and Decision Policies

The framework supports multiple configurable scenarios, each representing a different nurse behavior policy.

The active scenario is defined by two global parameters:

- GlobalSettings.methodToChooseRequest: specifies the strategy for task selection (e.g., CA trust model, FIFO);
- GlobalSettings.scenario:

Specifies high-level behavioral logic when nurses use the CA trust model:

1. Scenario 1 – Baseline: low-performing nurses limit their participation to tasks below a certain difficulty threshold ;
2. Scenario 2 – Replacement: low-performing nurses request substitution;
3. Scenario 3 – training with mentor: a trainer nurse assists the low-performing nurse, enabling performance improvement.

The modular scenario design allows easy extension for future behavioral studies.

3.1.5. Workflow

**Base Workflow**

At the start of the simulation, patients are spawned one by one and assigned to available beds following a predefined sequence (Bed 1->Bed 4->Bed 7->Bed 2->Bed 5->Bed 8->Bed 3->Bed 6-> Bed 9). This sequence ensures that all three doctors begin examining patients as early as possible. Once all beds are occupied, each subsequent patient is spawned only when a bed becomes free, immediately moving to occupy it and waiting to be examined by the corresponding doctor.

Each doctor leaves the staff waiting room as soon as the first patient lies down in one of their assigned beds. The doctor then moves next to the patient's bed, performs an examination (simulated with a fixed delay of 10 seconds), and then issues task requests to nurses (e.g., insertion of an intravenous (IV) catheter).

In the current implementation, the workflow models only one type of nursing task - IV catheter insertion. Extensions to include a broader range of nursing interventions are discussed in Future Work Section.

Nurses evaluate incoming requests based on their assigned decision-making policy and decide whether to execute or ignore them. When nurses accept a task, they move to the assigned patient, verify whether the task has already been completed by another nurse, and execute it if still pending. Nurses do not currently coordinate with each other to share or delegate tasks; however, such collaborative behavior could be added in future scenarios.

Task execution is simulated with a time delay, during which a floating UI label above the nurse displays the remaining time. Once the task is complete, the nurse returns to the staff waiting room to prepare for the next task (e.g., washing hands, gathering necessary materials). While these preparatory activities are not yet visually represented in the simulation, they could be incorporated in future iterations to enhance realism.

Nurses remain idle in the waiting room either when no tasks are available, or when they have self-assessed as low-performing and no tasks suitable to their skill level exist.

**Replacement Scenario Workflow**

In Scenario 2 (Replacement), nurses self-classified as low-performing using the CA trust model remain active in patient care but restrict their participation to low-difficulty tasks as in the Baseline scenario. To prevent delays in executing these higher-difficulty tasks, the simulation spawns an additional high-performing nurse, who takes over the pending requests of the low-performing nurse. The replacement nurse starts from the staff waiting room and begins executing these tasks immediately.

This approach models a hospital policy in which underperforming staff are supplemented with additional personnel rather than replaced entirely, ensuring that difficult tasks are still addressed promptly while allowing the original nurse to continue contributing within their capability range.

**Training Scenario Workflow**

In Scenario 3 (training with mentor), when a nurse determines her performance is low-performing, she requests the assistance of a dedicated trainer nurse. This trainer is spawned near the underperforming nurse and follows her to observe the execution of tasks. The trainer's role is to mentor the nurse through live task execution, improving her performance over time.

During the training phase, the trainee's task performance gradually improves based on observed tasks, with each completed observation incrementing the learning progress. The trainer directly supervises the execution process but does not perform the tasks themselves. Once the trainee's performance metrics surpass the required threshold, the training phase ends, the trainer is leaving ER, and the nurse resumes full responsibilities independently.

This scenario reflects real-world continuous knowledge transfer and skill improvement, where mentoring is provided as an alternative to staff replacement, aiming to restore full performance capabilities.

3.1.6. Logging and Metrics

The simulation automatically tracks and logs detailed performance metrics per shift, including:

- Number of patients served – Measures the total number of patients who have been fully attended to during the shift. A patient is considered "served" when all tasks requested by their assigned doctor have been successfully performed by the nurses. In addition to the total shift value, this metric is also recorded per doctor, enabling analysis of how the doctor's evaluation style (estimates correctly, overestimates, underestimates) relates to the number of patients served under their care;
- Total time damage – Quantifies the cumulative extra time taken by nurses to complete tasks beyond the expected task duration. This excess time is considered detrimental to patient outcomes – either by reducing the patient's survival probability (e.g., in cases such as cardiac arrest where an IV catheter must be placed quickly to administer life-saving medication) or by increasing patient dissatisfaction due to prolonged waits. The simulation computes this metric by summing all positive deviations (actual duration – expected duration) for completed tasks. In addition to the overall value for each shift, time damage is recorded per nurse – enabling analysis of how a nurse's quality attribute (e.g., high-performing, low-performing) correlates with the delays they cause – and per doctor, where it reflects the cumulative delays affecting that doctor's patients. This doctor-level recording allows investigation of how a doctor's evaluation style (estimates correctly, overestimates, underestimates) influences the delays, e.g., whether underestimation of task difficulty leads to higher time damage;
- Total delay – Measures the total amount of time patients remain unattended by any nurse after they have been examined by their assigned doctor and one or more tasks requests have been sent to nurses. These are pure waiting periods in which the patient occupies a bed but receives no active nursing intervention. In addition to the overall shift value, total delay is also recorded per doctor, allowing the analysis of how each doctor's evaluation style influences the waiting times experienced by their patients;
- Task success/failure counts – For each nurse, the simulation records the number of successfully and unsuccessfully completed tasks. A task is counted as successful if the actual execution time does not exceed the time estimated and requested by the doctor who examined the patient. These metrics therefore represent the doctor's assessment of each nurse's performance in executing assigned tasks. As a result, due to a doctor's systematic bias in evaluation style (see Section 3.1.2), a task may be labeled unsuccessful despite meeting the gold-standard duration (in cases of overestimation), or labeled successful despite exceeding it (in cases of underestimation);
- Utility values for each nurse – For each nurse, the simulation records a cumulative utility score based on successfully completed tasks. When a task is successful (i.e., its actual execution time does not exceed the time estimated and requested by the examining doctor), the nurse's utility increases by the task's requestedLevel – an integer from 1 (easiest) to 5 (most difficult) representing the task's difficulty. This metric can serve as a motivational indicator, with nurses potentially aiming to maximize their utility over the course of the shift;
- Doctors delays and evaluation accuracy.

These metrics are displayed in the in-game UI and exported to external .csv files for post-simulation statistical analysis.

3.1.7. Customizability and Reproducibility

The platform is designed for experimental flexibility:

- Agent counts, roles, and attributes are configured via the GlobalSettings class;
- Scenario-specific logic is encapsulated for modular extension;
- Run IDs and parameter logging ensure reproducibility of results.

This modular, data-driven design supports iterative development of new scenarios and facilitates rigorous comparative analysis.

*3.2. Decision-Support Experimentation Methodology*

3.2.1. Objectives and Rationale

The primary aim of the decision-support experiments is to evaluate how alternative nurse decision-making policies and workflow strategies influence key operational outcomes in a simulated emergency room (ER) environment. By systematically varying agent behaviors and resource allocation strategies, the framework allows us to identify policies that minimize patient delays, reduce detrimental time damage, and optimize the utilization of available staff.

This approach supports data-driven decision making in hospital management, enabling stakeholders to test "what-if" scenarios before implementing changes in real clinical settings.

3.2.2. Experimental Design

**Scenarios**

Three distinct operational scenarios within the simulation framework can be evaluated. Each represents a different policy for handling low-performing nurses.

Scenario 1 – Baseline: Low-performing nurses remain active but only perform tasks below a predefined difficulty threshold. More complex tasks are skipped by these nurses and remain unassigned until another available nurse chooses to execute them. Unlike the Replacement scenario, where an additional high-performing nurse assists with complex tasks, and the Training scenario, where a trainer nurse accompanies and guides the low-performing nurse, no extra nursing staff is provided in the Baseline scenario.

Scenario 2 – Replacement: Low-performing nurses remain active but only perform tasks below a certain difficulty threshold; and additional high-performing nurse is spawned to execute pending higher-difficulty tasks, ensuring they are addressed without excessive delay.

Scenario 3 – Training with Mentor: Low performing nurses are accompanied by a trainer nurse who supervises and guides them, improving their performance over time.

It is important to note that the FIFO decision-making policy is only meaningful in the Baseline scenario, as it does not support performance self-assessment or adaptive responses to low quality. In contrast, the CA trust model is applicable across all scenarios and is essential in Replacement and Training settings, as it enables nurses to detect low performance and trigger the corresponding assistance or mentoring mechanisms.

**Controlled Variables**:

- Number of doctors, and nurses;
- ER layout and bed allocation rules;
- Patient assignment sequence – fixed order in which patients occupy available beds to maximize early doctor engagement (1->4->7->2->…);
- Task types;
- Evaluation styles of doctors and quality attributes of nurses (unless explicitly varied).

**Independent Variables**:

- Nurse decision-making policy (CA trust model, FIFO);
- Scenario type (Baseline, Replacement, Training);
- Doctor evaluation styles (overestimates, underestimates, estimates correctly);
- Nurse quality attributes (high-performing, low-performing).

**Dependent Variables**:

The simulation tracks a set of predefined performance metrics (see Section Logging and Metrics), including:

- Number of patients served;
- Total delay;
- Total time damage;
- Task success/failure counts;
- Utility values per nurse.

3.2.3. Experimental Procedure

The experiments are conducted for four specific scenario-policy combinations:

1. Baseline scenario – CA trust model;
2. Baseline scenario - FIFO;

3. Replacement scenario – CA trust model;
4. Training scenario – CA trust model.

For each of these combinations, the simulation is executed 60 times to account for stochastic variations in task durations and agent behaviors, with different random seeds to ensure statistical robustness. In the Replacement and Training scenarios, only the CA trust model is tested, as this decision-making policy is the only one that allows nurses to detect their low performance and either request assistance from an additional high-performing nurse (Replacement) or receive guidance from a trainer nurse (Training). The FIFO policy is applied only in the Baseline scenario.

Simulation parameters are initialized via the GlobalSettings class. A unique Run ID is assigned to each execution for reproducibility. Patient spawning, doctors' examinations, and nurse tasks executions proceed according to the workflows described in Section Workflow. Performance metrics are automatically logged both in the in-game UI and in external CSV files.

3.2.4. Data Analysis Approach

Collected data are analyzed using descriptive statistics and inferential statistical tests (e.g., ANOVA, t-tests, or non-parametric equivalents where appropriate) to determine whether observed differences between policies are statistically significant.

From a decision-support perspective, the results highlight trade-offs between different strategies. For instance, the Replacement scenario may significantly reduce total delay for complex tasks but could require additional staffing resources, while the Training scenario may offer long-term benefits by improving low-performing nurses without permanently increasing headcount.

This analysis framework enables hospital managers and policymakers to:

1. Predict the operational impact of specific interventions before real-world deployment;
2. Compare alternative policies under controlled and repeatable conditions;
3. Identify scenario-specific strengths and weaknesses, informing adaptive staffing and training strategies.

## 4. Results

*4.1. Case Study Example*

To illustrate the practical use of the simulation framework for decision support, we present a case study of a small emergency room (ER) setup. The configuration includes three doctors (all accurately estimating task requests), two nurses (one high-performing and one low-performing), and nine beds. A detailed description of the experimental setup – including objectives and rationale, statistical analysis, results and conclusions – is provided in Appendix B (Decision Support Case Study). In this section, we focus only on the key findings, their interpretation, and the main limitations, along with the conclusions drawn.

*4.2. Key Findings*

**Baseline Scenario**: The CA trust model prioritizes patient safety by detecting low-performing nurses and limiting their engagement with complex tasks. This reduces the risk of errors but may lead to longer patient delays. The FIFO method, in contrast, improves fairness and reduces delays, yet increases exposure to potential nurse-related errors. This demonstrates the trade-off between safety and operational efficiency.

**Replacement Scenario**: Adding an extra high-performing nurse to handle complex tasks significantly improves patient throughput and reduces delays. The immediate operational gains are clear; however, they come at the cost of additional staffing, with associated financial and logistical implications.

**Training Scenario**: Pairing low-performing nurses with a trainer nurse improves their skill over time, increasing task completion rates and stabilizing high-performing nurses' workload. Although delays and patient damage may be higher in the short term, this scenario fosters long-term capacity building.

**Replacement vs. Training**: Replacement yields immediate operational benefits, including faster patient service and reduced cumulative delays, whereas Training enhances the capabilities of low-performing nurses over time. Decision-makers must weigh the short-term efficiency gains of Replacement against the long-term development benefits of Training, particularly under constrained staffing resources.

*4.3. Interpretation and Limitations*

This case study is based on a specific configuration of doctors, nurses, and beds. Results may vary significantly under different conditions – for example, if doctor evaluation styles differ (overestimating or underestimating tasks) or if the nurse composition changes (e.g., two high-performing nurses). Additionally, the type of task being simulated can strongly influence outcomes. In this study, IV catheter insertion was used – a task where improvement of low-performing nurses occurs gradually over time in the Training scenario. For other tasks, such as vital sign measurement, improvement may occur rapidly. In such cases, even a single demonstration might suffice for a low-performing nurse to reach a high performance, potentially making the Training scenario more effective than Replacement across all performance indicators for that task. Therefore, the findings should be interpreted as illustrative of the framework's capabilities rather than universally generalizable conclusions.

*4.4. Conclusions*

The case study highlights how the simulation framework supports evidence-based decision making by allowing hospital managers to explore trade-offs between patient safety, operational efficiency, and staff development under controlled, repeatable conditions. This demonstrates that policy selection depends on strategic priorities, and highlights the potential of the framework as a decision-support tool to test alternative scenarios before real-world implementation.

## 5. Discussion

The results of our simulation experiments demonstrate the potential of trust-based allocation mechanisms to enhance decision-making in emergency department (ED) staffing and workflow coordination. By comparing baseline, replacement and training scenarios under both FIFO and CA trust-based decision policies, our study highlights key trade-offs between patient safety, operational efficiency, and staff development.

A central finding is that the CA trust model consistently prioritizes patient safety by detecting low-performing nurses and limiting their exposure to high-complexity tasks. This reduces the likelihood of critical errors but introduces longer delays, especially in the baseline scenario. In contrast, the FIFO policy improves fairness and reduces waiting times but exposes patients to greater risks associated with incorrect or incomplete task execution. This contrast underscores the inherent tension between safety-oriented and efficiency-oriented policies in high-pressure ED environments.

The introduction of replacement and training interventions provides complementary strategies for mitigating the limitations of the baseline configuration. The replacement scenario offers immediate operational benefits by assigning high-complexity tasks to an additional high-performing nurse, thereby reducing both patient delays and cumulative time damage. However, these gains come with increased resource demands and higher operational costs, which may not be sustainable under staffing or budgetary constraints. The training scenario, by contrast, provides a mechanism for capacity building: low-performing nurses gradually improve through guided practice with a trainer nurse. While this incurs short-term delays and higher patient risk, it fosters long-term workforce development and reduces reliance on permanent staffing increases. Together, these findings suggest that replacement and training are not mutually exclusive but may be strategically combined depending on institutional priorities and resource availability.

From a broader perspective, our findings contribute to the growing literature on multi-agent trust and reputation systems by demonstrating their applicability to healthcare operations. While trust-based models have been widely studied in domains such as online marketplaces, peer-to-peer systems, and the Internet of Things, their integration into healthcare workforce management remains underexplored. By modeling nurses and doctors as intelligent agents capable of self-assessment and adaptive decision-making, our framework extends computational trust research into a domain where the stakes of decision errors are exceptionally high.

The case study analysis further illustrates how the framework can support hospital managers in evidence-based decision-making. For example, a small ED with limited staff may initially favor replacement strategies to ensure patient throughput, but transition to training strategies once capacity constraints are stabilized. By offering a sandbox for "what-if" experimentation, the framework allows decision-makers to explore the consequences of operational policies before implementation, reducing the risks of trial-and-error approaches in real clinical settings.

Nevertheless, the study has several limitations that must be acknowledged. First, the results are derived from a restricted configuration of doctors, nurses, and tasks, limiting their direct generalizability. Second, the current implementation models only a single nursing task (IV catheter insertion), which does not fully capture the diversity of ED workflows. Finally, agent behaviors are rule-based, meaning that contextual variability, learning effects, and human factors such as fatigue or stress are not yet represented. These constraints suggest that the findings should be interpreted as illustrative of the framework's capabilities rather than prescriptive for real-world decision-making.

Looking ahead, the framework provides a foundation for more adaptive and realistic modeling of ED operations. As outlined in the Future Work Section, integrating data-driven behavioral models, customizing ED environments to reflect site-specific layouts, and expanding the range of nursing tasks will significantly increase ecological validity. Such enhancements will strengthen the framework's role as a practical decision-support tool, enabling hospital managers to balance short-term efficiency with long-term workforce sustainability.

In summary, the present work highlights the importance of trust-based allocation mechanisms in managing uncertainty and variability in emergency care. By explicitly modeling the trade-offs between safety, efficiency, and staff development, our framework offers both theoretical insights into computational trust models and practical decision-support value for healthcare institutions.

## 6. Conclusions and Future Work

This work presented a simulation-based framework that leverages computational trust models to support decision-making in emergency department (ED) staffing and workflow coordination. By modeling doctors and nurses as intelligent agents within a Unity-based simulation, we demonstrated how trust-based task allocation mechanisms can influence team performance, patient safety, and resource utilization. Unlike traditional rule-based simulations, our framework emphasizes the dynamics of interpersonal trust, offering a more realistic representation of clinical decision-making processes. Experiments across alternative policies (Baseline, Replacement, Training) illustrated how trust-aware coordination can reduce patient delays, mitigate errors, and highlight staff training needs, providing valuable insights for evidence-based healthcare management. A further contribution lies in linking trust-driven coordination with the principles of personalized medicine, showing how adaptive allocation can evolve toward patient-specific care.

Having demonstrated the applicability of our simulation framework for modeling staff coordination and trust-based task allocation in an emergency department (ED) setting, we envision several directions for future research. These directions aim to increase the realism of agent behavior, to better capture the variability of real-world emergency workflows, and to expand the decision-support capabilities of the system. By moving beyond fixed, rule-based representations, the framework can evolve into a more adaptive platform capable of supporting evidence-based policy evaluation and operational planning.

*6.1. Learning Agent Behavior Models from Logged Performance Data*

In the current implementation, both doctor evaluation styles and nurse quality attributes are modeled through fixed, rule-based functions (see Section 3.1.2). While this design enables controlled comparisons between scenarios, it does not capture the variability and context-dependent decision-making that characterize real ED staff.

A promising extension is to replace these hard-coded rules with data-driven models inferred from the simulation's own detailed performance logs or from real-world clinical data.

6.1.1. Inferring Doctor Evaluation Styles

Doctor-level logs include total patient time damage, cumulative delays before treatment, and the number of patients served. Analyzing the relationships among these metrics could allow inference of doctor evaluation styles. For instance, systematic underestimation of task difficulty may correlate with increased patient damage, while overestimation may inflate perceived nurse error rates. Machine learning methods such as classification, clustering, or Bayesian inference could estimate probability distributions over evaluation styles, leading to more realistic physician-agent behaviors.

6.1.2. Inferring Nurse Quality Attributes

Nurse-level logs record task outcomes (success/failure counts), time damage, delay impacts, and utility values. While the current framework categorizes nurses as simply high-performing or low-performing, this binary classification oversimplifies reality. Regression models, clustering approaches, or reinforcement learning could be employed to infer richer nurse skill profiles, capturing variability due to experience, fatigue, training, or workload.

By adopting this dual-level learning approach, the framework could evolve into a self-adaptive platform in which agent parameters are continuously updated based on observed behavior, supporting more accurate predictive modeling and more credible decision-support analyses.

*6.2. Environment and Workflow Customization*

Currently, the ED environment is modeled with a fixed number of beds, a predefined spatial layout, and static doctor-to-bed assignments (e.g., Doctor 1 handles Beds 1-3, etc.). While this design ensures reproducibility, it does not reflect the diversity of layouts and operational policies found in real-world EDs.

Future extension could include dynamic generation of ED layouts based on hospital blueprints, diagrams, or even digital twins of existing facilities. This would enable replication of site-specific infrastructural constraints such as bed capacity, aisle widths, nurse stations, and patient flow paths.

Moreover, the framework could incorporate hospital-specific operational rules, such as custom doctor-to-bed mappings, shift rotations, or nurse allocation policies. Such features would allow testing of interventions tailored to individual institutions, thereby enhancing the practical decision-support value of the simulation.

*6.3. Expanding the Range of Nursing Tasks*

At present, the framework models only one type of nursing intervention – IV catheter insertion. While useful for demonstrating the core dynamics of trust-based allocation, this narrow focus limits the generalizability of conclusions.

Future work could extend the simulation to include a wider variety of common ED tasks such as medication administration, blood pressure measurement, and electrocardiograms (ECG). Expanding the task set would diversify nurse decision-making contests and allow systematic exploration of task prioritization strategies. Importantly, it would also capture different skill acquisition trajectories: for example, certain tasks (e.g., vital signs measurement) may allow rapid performance improvement after limited training, while others (e.g., IV catheter insertion) require gradual skill development.

This richer task environment would not only increase the ecological validity of the simulation but also enable more nuanced evaluation of staff management strategies. In particular, it would allow testing whether the relative benefits of Replacement versus Training policies vary across task types – a key insight for hospital managers aiming to balance short-term efficiency with long-term staff development.

This study highlights the potential of computational trust models to transform how we understand and optimize emergency department operations. By integrating trust-driven coordination into agent-based simulations, we bridge the gap between individual decision-making, team performance, and patient-centered outcomes. Looking ahead, trust aware mechanisms could form the basis for adaptive, AI-driven systems that not only improve efficiency but also support resilient and personalized models of care, positioning this work as a stepping stone toward the next generation of healthcare decision-support tools.

## Appendix A. Algorithms

*Appendix A.1.*

**Table A1.** Doctor's performance level estimation based on evaluation style.

**Algorithm 1:** Doctor's performance level estimation based on evaluation style

```
function EvaluatePerformanceLevel(trueLevel, evaluationStyle):
   switch(evaluationStyle):

      case "overestimates":
         if trueLevel <= 2: return 3
         if trueLevel >= 3: return 5

      case "underestimates":
         if trueLevel >= 4: return 3
         if trueLevel <= 3: return 1

      case "estimatesCorrectly":
         return trueLevel

      default:
         return trueLevel
```

*Appendix A.2.*

**Table A2. Nurse task execution time determination based on quality and training status**.

**Algorithm 2:** Nurse task execution time determination based on quality and training status

```
function GetTaskDuration(nurseId, truePerformanceLevel, observedTasks):
   roll = random(0, 1)

   if nurseId == 1:  // Low-performing nurse
      isTrainingScenario = (methodToChooseRequest == 1 AND Scenario == 3)

      if isTrainingScenario:
         bonusChance = clamp(observedTasks * 0.1, 0, 1) // each observed task adds 10%

         if roll <= bonusChance:
            return baseDurationForLevel(truePerformanceLevel)
         else:
            return randomDurationAboveBase(truePerformanceLevel)
      else:
         return randomDurationAboveBase(truePerformanceLevel)

   else if nurseId == 2: // High-performing nurse
      rollGood = random(0, 1)

      if rollGood <= 0.90:
         return baseDurationForLevel(truePerformanceLevel)
      else:
         return randomDurationAboveBase(truePerformanceLevel)

   return 40 // default fallback
```

```
function baseDurationForLevel(level):
    switch(level):
        case 1: return 60
        case 2: return 50
        case 3: return 40
        case 4: return 30
        case 5: return 20
        default: return 40

function randomDurationAboveBase(level):
    // returns a duration in a higher range to simulate delay
    switch(level):
        case 1: return random(60, 70)
        case 2: return random(50, 70)
        case 3: return random(40, 50)
        case 4: return random(30, 40)
        case 5: return random(20, 30)
        default: return 40
```

## Appendix B. Decision Support Case Study

*Appendix B.1. Comparative Evaluation of Nurse Task Selection Methods in the Baseline Scenario*

**Objective**

This case study investigates how two alternative nurse task selection policies affect patient care quality and operational performance metrics in the Baseline scenario:

- Method 1: CA trust model;
- Method 2: FIFO.

**Design**

- Scenario: Baseline (no replacement nurses, no training interventions);
- Task selection policies: CA trust model vs. FIFO;
- Simulation runs: 60 independent runs per method;
- Metrics analyzed: Total patients served, total patient damage time, total patient delay, failed and successful tasks per nurse, nurse utility, nurse time damage, patient distribution across doctors.

**Statistical Approach**

All metrics were tested for normality using the Shapiro-Wilk test. Due to significant deviations from normality in most cases (p<0.001), non-parametric Wilcoxon rank-sum tests were applied unless otherwise specified. Effect sizes, medians, means and p-values are reported for each comparison.

**Key Results**

Total Patients Served: Since both groups exhibited significant deviations from normality (Shapiro Wilk p<0.001), the non-parametric Wilcoxon rank-sum test was applied. The analysis showed that the FIFO policy served significantly more patients than the CA trust model ($W = 0, p < 2.2 \times 10^{-16}$). The mean difference was approximately 11 patients per shift (CA trust: 22.18, FIFO: 33.22), indicating that prioritizing tasks by arrival order rather than trust scores substantially increases throughput in the Baseline scenario.

Total Patient Damage Time: Both groups deviated significantly from normality (Shapiro-Wilk p<0.001), so the non-parametric Wilcoxon rank-sum test was applied. The CA trust model resulted in dramatically lower cumulative patient damage time compared to the FIFO policy ($W = 0, p < 2.2 \times 10^{-16}$). On average, the damage time was reduced by nearly 200 seconds per shift (CA trust: 65.38 s, FIFO: 262.77 s), indicating that trust-based task allocation substantially mitigates patient exposure to harmful delays in treatment.

Total Patient Delay: Both groups showed strong deviations from normality (Shapiro-Wilk p<0.001), so the Wilcoxon rank-sum test was used. FIFO produced significantly lower total patient delay compared to the CA trust model ($W = 90000, p < 2.2 \times 10^{-16}$). The mean reduction was about 642 seconds per shift (CA trust: 7095.77 s, FIFO: 6453.23 s), suggesting that FIFO's task allocation leads to shorter waiting periods for patients before nursing intervention. However, as earlier results show, this comes at the cost of increased patient damage time.

Failed Tasks by the Low-performing Nurse: Both groups exhibited strong deviations from normality (Shapiro-Wilk, p<0.001), so the non-parametric Wilcoxon rank-sum test was applied. Results showed that the low-performing nurse failed significantly more tasks under FIFO (Method 2) than under the CA trust model (Method 1) ($W = 3600, p < 2.2 \times 10^{-16}$). The mean failure count was 13.53 for Method 2 and 3.60 for Method 1, a difference of nearly +10 failed tasks per shift. This confirms that FIFO's inability to avoid high-difficulty tasks for low-performing nurses leads to a substantial increase in failed task executions.

Failed Tasks by the High-performing Nurse: Normality tests indicated significant deviations from the normal distribution in both groups (Shapiro-Wilk p<0.001), so the Wilcoxon rank-sum test was used. The test did not reveal a statistical significant difference in failed tasks between FIFO and the CA trust model ($W = 1852.5, p = 0.389$). Mean failure counts were very similar (Method 1: 1.92, Method 2: 1.98), and the distributions showed substantial overlap. These results suggest that high-performing nurses maintain a low and stable error rate regardless of the task allocation policy.

Successful Tasks by the Low-performing Nurse: Both groups contained only zero values for the number of successful tasks, making normality testing infeasible and rendering statistical comparisons trivial. The Wilcoxon rank-sum test confirmed no difference between Method 1 (CA trust model) and Method 2 (FIFO) ($W = 1800, p = 1$). In other words, low-performing nurses were unable to complete any tasks successfully under either task allocation policy, highlighting that policy choice alone does not improve task success when nurse performance is intrinsically poor.

Successful Tasks by the High-performing Nurse: Shapiro-Wilk tests indicated that Method 1 data significantly deviated from normality (p<0.001), while Method 2 data showed no significant deviation (p=0.075). Given the distributional mismatch, the non-parametric Wilcoxon rank-sum test was applied. Results showed no significant difference in the number of successful tasks between Method 1 (median = 17) and Method 2 (median = 18) ($W = 1853, p = 0.391$). These findings suggest that highly skilled nurses maintain similar success rates regardless of whether patient allocation follows the CA trust model or a FIFO policy.

Nurse Utility by the Low-performing Nurse: Shapiro-Wilk tests indicated that Method 1 data significantly deviated from normality (p=0.006), while Method 2 data showed no significant deviation (p=0.517). Given the distributional mismatch, the non-parametric Wilcoxon rank-sum test was applied. Results showed that the low-performing nurse had significantly lower utility in Method 2 (median = -44.50) compared to Method 1 (median = -9.97) ($W = 0, p < 2.2 \times 10^{-16}$). These findings suggest that the low-performing nurse performs substantially worse under Method 2 conditions.

Nurse Utility by the High-performing Nurse: Shapiro-Wilk tests indicated that Method 1 data significantly deviated from normality (p=0.027), while Method 2 data showed no significant deviation (p=0.916). Given the distributional mismatch, the non-parametric Wilcoxon rank-sum test was applied. Results showed no significant difference in nurse utility between Method 1 (median = 50.28) and Method 2 (median = 53.25) (W=1749, p=0.607). These findings suggest that highly skilled nurses maintain similar utility regardless of whether patient allocation follows Method 1 or Method 2.

Nurse Time Damage by the Low-performing Nurse: Shapiro-Wilk tests indicated that Method 1 data significantly deviate from normality (p=0.002), while Method 2 data showed no significant deviation (p=0.582). Given the distributional mismatch, the non-parametric Wilcoxon rank-sum test was applied. Results showed a significant difference in NurseTimeDamage between Method 1 (median = 56.48) and Method 2 (median = 252.37) ($W = 0, p < 2.2 \times 10^{-16}$). These findings suggest that low-performing nurses cause substantially lower time damage when patient allocation follows Method 1 compared to Method 2.

Nurse Time Damage by the High-performing Nurse: Shapiro-Wilk tests indicated that NurseTimeDamage data for both Method 1 (p=0.003) and Method 2 (p=0.003) significantly deviated from normality. Given the non-normal distributions, the non-parametric Wilcoxon rank-sum test was applied. Results showed no significant difference in NurseTimeDamage between Method 1 (median = 8.87) and Method 2 (median = 10.32) (W=1679, p=0.263). These findings suggest that highly skilled nurses maintain similar time damage regardless of whether patient allocation follows Method 1 and Method 2.

Patient Distribution Across Doctors: Nurses prefer some doctors in Method 1, while patients are served uniformly in Method 2. Using the Monte Carlo $x^2$ test for each simulation run, in Method 1, 18 out of 60 runs showed a statistically significant nurse preference for certain doctors (p<0.05). In contrast, in Method 2, none of the runs showed a significant preference. The variance in the number of patients served per doctor per run was much higher in Method 1 (mean variance = 21.44) compared to Method 2 (mean variance = 0.6). A paired t-test confirmed that this difference is statistically significant $p \approx 1.46e-09$). The conclusion is that nurses appear to favor certain doctors in Method 1, resulting in an uneven distribution of patients. In Method 2, patients are distributed uniformly across doctors, as expected.

### Conclusions

Method 1 (CA trust model) prioritizes patient safety and reduces the risk from low-performing nurses but can result in longer patient delays and potential doctor preference bias.

Method 2 (FIFO) improves fairness and reduces delays but increases exposure to nurse-related damage.

These results highlight a trade-off between safety and operational efficiency. Hospitals can select the appropriate method depending on their priorities: safety-focused institutions may prefer Method 1, while those emphasizing throughput and fairness may prefer Method 2.

*Appendix B.2. Baseline (CA trust model) vs. Replacement (CA trust model)*

### Objective

Evaluate the impact of introducing a high-performing replacement nurse for tasks exceeding the difficulty threshold of low-performing nurses.

### Design

- Scenarios: Baseline vs. Replacement;
- Task selection policy: CA trust model (applied in both scenarios);
- Metrics analyzed: Total patients served, total patient damage time, total patient delay, failed and successful tasks per nurse, nurse utility, nurse time damage, patient distribution across doctors.

### Statistical Approach

All metrics were tested for normality using the Shapiro-Wilk test. Due to significant deviations from normality in most cases, non-parametric Wilcoxon rank-sum tests were applied. For metrics with no variance, no statistical test was meaningful. Effect sizes, medians, means, and p-values are reported for each comparison.

### Key Results

Total Patients Served: Since both groups exhibited significant deviations from normality (Shapiro-Wilk p<2.2e-16), the non-parametric Wilcoxon rank-sum test was applied. Results showed that the Replacement scenario served significantly more patients ($W = 1935, p < 2.2 \times 10^{-16}$) compared to the baseline. The mean difference was approximately 19.4 patients per shift (Baseline: 22.18, Replacement: 41.58), indicating that adding an extra nurse to handle high-difficulty tasks substantially increases the overall patient throughput.

Total Patient Damage Time: Since both groups showed significant deviations from normality (Shapiro-Wilk test: Baseline $p = 7.29 \times 10^{-9}$, Replacement $p = 1.63 \times 10^{-4}$, the non-parametric Wilcoxon rank-sum test was applied. Results revealed that the Replacement scenario had a significantly higher total time in damage (W=27360, $p < 2.2 \times 10^{-16}$) compared to the Baseline. The mean difference was approximately 30.6 seconds per shift (Baseline: 65.38 sec, Replacement: 95.97 sec), indicating that the staffing change, while improving throughput, also increased the cumulative time that patients spent in a critical condition.

Total Patient Delay: Shapiro-Wilk tests indicated that both the Baseline (CA trust) data (W=0.787, p<0.001) and the Replacement (CA trust) data (W=0.652, p<0.001) significantly deviated from normality. Given the non-normal distributions, the non-parametric Wilcoxon rank-sum test was applied, revealing a significant difference in total patient delay between the two scenarios ($W = 106605, p < 0.001$). Specifically, the Replacement scenario exhibited a substantially lower mean total delay (M=6284.12 s) compared to the Baseline scenario (M=7095.77 s) representing an average reduction of approximately 811.65 seconds per shift. These findings suggest that adding an extra nurse to handle high-difficulty tasks significantly decreases overall patient delay in the emergency room.

Failed Tasks by the Low-performing nurse: Since both groups exhibited significant deviations from normality (Shapiro-Wilk p<0.001), the non-parametric Wilcoxon rank-sum test was applied. Results indicated a small difference in the number of failed tasks between the Baseline and Replacement scenarios (W=1416.5, p=0.037), with the mean number of failed tasks being slightly higher in the Replacement scenario (Baseline: 3.6, Replacement: 4.0). This suggests that adding an extra nurse for high-difficulty tasks did not substantially reduce failed tasks, and the practical difference between the two scenarios is minimal.

Failed Tasks by the High-performing nurse: Since both groups exhibited significant deviations from normality (Shapiro-Wilk p<0.001), the non-parametric Wilcoxon rank-sum test was applied. Results showed no statistically significant difference in the number of failed tasks between the Baseline and the Replacement scenario (W=1744.5, p=0.7642). The mean number of failed tasks was identical across both scenarios (Baseline: 1.92, Replacement: 1.92), indicating that replacing one nurse did not affect the failure rate of high-performing nurse.

Successful Tasks by the Low-performing nurse: Since both groups exhibited no variance (all values were identical at 0), neither the Shapiro-Wilk test for normality nor inferential statistical tests (Wilcoxon, Welch's t-test) could be meaningfully applied. Both the Baseline and Replacement scenarios resulted in a mean of 0 successful tasks per shift, indicating that the low-performing nurse was unable to complete any successful tasks under either condition.

Successful tasks by the High-performing nurse: Since both groups exhibited significant deviations from normality (Shapiro-Wilk p<0.001), the non-parametric Wilcoxon rank-sum test was applied. Results showed no significant difference between the Baseline and Replacement scenarios (W=1566.5, p=0.218). The mean difference was approximately 1.25 successful tasks (Baseline: 16.67, Replacement: 17.92), indicating that the high-performing nurse achieved a comparable number of successful tasks in both conditions.

Nurse Utility by the Low-performing nurse: Since both groups exhibited significant deviations from normality (Shapiro-Wilk p=0.006 for Baseline, p = 0.0009 for Replacement), the non-parametric Wilcoxon rank-sum test was applied. Results showed that the Replacement scenario did not differ significantly in Nurse Utility of the low-performing nurse compared to the Baseline (W= 1606, p=0.154). The mean Nurse Utility was slightly lower for Replacement (-10.47) than for Baseline (-9.97), with a mean difference of approximately 0.5 units per nurse, indicating that replacing the low-performing nurse in Scenario 2 did not substantially change the overall nurse utility under Method 1.

Nurse Utility by the High-performing Nurse: Since both groups showed deviations from normality (Shapiro-Wilk p<0.05 for both Baseline and Replacement), the non-parametric Wilcoxon rank-sum test was applied. Results indicated that the Replacement scenario had significantly higher nurse utility (W=2129.5, p=0.042) compared to the Baseline. The mean difference was approximately 6.17 units of utility (Baseline: 50.28, Replacement: 56.45) suggesting that the Replacement scenario improves the efficiency of the high-performing nurse in Method 1.

Nurse Time Damage by the Low-performing Nurse: Since both groups exhibited significant deviations from normality (Shapiro-Wilk test, Baseline: W=0.930, p=0.0021, Replacement: W=0.938, p=0.0045), the non-parametric Wilcoxon rank-sum test was applied. Results showed no significant difference between Baseline and Replacement (W=1955.5, p=0.794), indicating that the time damage caused by the low-performing nurse remained similar across scenarios. The mean NurseTimeDamage was slightly higher in the Replacement scenario (Baseline: 56.48, Replacement: 58.07), but this difference was not statistically significant. Thus, replacing the nurse team configuration did not reduce the time damage caused by the low-performing nurse.

Nurse Time Damage by the High-performing Nurse: Since both groups exhibited significant deviations from normality (Shapiro-Wilk p<0.05), the non-parametric Wilcoxon rank-sum test was applied. Results indicated no significant difference in Nurse Time Damage between the Baseline and Replacement scenarios (W=1853, p=0.611). The mean values were highly comparable (Baseline: 8.87, Replacement: 9.22), suggesting that introducing an additional nurse in the Replacement scenario does not substantially affect the workload-related time damage for the high-performing nurse.

Patient Distribution Across Doctors: In the Baseline scenario (Scenario 1), some nurses show a preference for certain doctors, while in the Replacement scenario (Scenario 2), the distribution is slightly more even. Using a Monte Carlo $x^2$ test for each simulation run, 18 out of 60 runs in the Baseline scenario showed a statistically significant nurse preference for specific doctors (p<0.05). In the Replacement scenario, 12 out of 60 runs showed a significant preference.

The variance in the number of patients served per doctor per run was 21.44 for the Baseline scenario and 29.71 for the Replacement scenario. A paired t-test comparing these variances found no statistically significant difference (p=0.938).

Conclusion: Nurses occasionally favor certain doctors in both scenarios, but the evidence is not strong enough to indicate a systematic preference. Patient distribution across doctors is generally uneven in both Baseline and Replacement scenarios, with no significant difference in variance between them.

**Conclusions**

The Replacement scenario, which adds an extra nurse to handle high-difficulty tasks, substantially increases overall patient throughput and reduces patient delay. However, these operational gains come with the implicit cost of employing additional staff. While total patient damage time rises slightly, the main trade-off to consider is the financial and logistical burden of the extra nurse. Decision-makers should weigh the benefits of higher efficiency and reduced patient delays against the increased staffing costs, particularly in resource-constrained settings.

*Appendix B.3. Baseline (CA trust model) vs. Training (CA trust model)*

**Objective**

Evaluate the impact of assigning trainer nurses to accompany low-performing nurses within the CA Trust model. The focus is on whether this intervention improves overall system throughput and efficiency, while accounting for potential trade-offs in error rates, patient safety, and nurse performance outcomes.

**Design**

- Scenarios: Baseline vs. Training;
- Task selection policy: CA trust model applied in both scenarios;
- Metrics analyzed: Total patients served, total patient damage time, total patient delay, failed and successful tasks per nurse, nurse utility, nurse time damage, patient distribution across doctors.

**Statistical Approach**

All metrics were tested for normality using the Shapiro-Wilk test. Given that most metrics significantly deviated from normality, the non-parametric Wilcoxon rank-sum test was applied. For metrics with no variance in one group, the test was applied against the non-normal counterpart. Effect sizes, means, and p-values are reported. For patient distribution, Monte Carlo chi-square tests and variance comparisons via paired t-tests were conducted.

**Key Results**

Total Patients Served: Since both groups exhibited significant deviations from normality (Shapiro-Wilk $p < 2.2 \times 10^{-16}$), the non-parametric Wilcoxon rank-sum test was applied. Results showed that the Training scenario served significantly more patients ($W = 10850, p < 2.2 \times 10^{-16}$) compared to the baseline. The mean difference was approximately 6.93 patients per shift (Baseline: 22.18, Training: 29.12), indicating that assigning trainer nurses to accompany low-performing nurses substantially increases overall patient throughput.

Total Patient Damage Time: Since both groups deviated significantly from normality (Shapiro-Wilk $p < 2.2 \times 10^{-16}$ for Baseline, $p = 6.73 \times 10^{-5}$ for Training), the non-parametric Wilcoxon rank-sum test was applied. Results showed that the Training scenario had significantly higher total patient damage time ($W = 7025, p < 2.2 \times 10^{-16}$) compared to the Baseline. The mean difference was approximately 83.9 seconds per shift (Baseline: 65.38 sec, Training: 149.28 sec), indicating that assigning trainer nurses increased the overall patient damage time.

Total Patient Delay: Since both groups exhibited significant deviations from normality (Shapiro-Wilk $p < 2.2 \times 10^{-16}$), the non-parametric Wilcoxon rank-sum test was applied. Results showed that the Training scenario had significantly lower total patient delay (W=83100, $p < 2.2 \times 10^{-16}$) compared to the Baseline. The mean difference was approximately 380.9 seconds per shift (Baseline: 7095.77 s, Training: 6714.88 s), indicating that implementing the training scenario reduces patient delays and potentially improves overall efficiency in the emergency department.

Failed Tasks by the Low-performing nurse: Since failures in the Baseline group exhibited significant deviations from normality (Shapiro-Wilk W=0.914, p=0.00045), while the Training group did not (W=0.967, p=0.102), the non-parametric Wilcoxon rank-sum test was applied. Results showed that the low-performing nurse committed significantly more failed tasks in the Training scenario ($W = 140, p < 2.2 \times 10^{-16}$) compared to the Baseline. The mean difference was approximately 4.28 failed tasks per shift (Baseline: 3.60, Training: 7.88). This outcome suggests that, although trainer nurses improve overall system throughput, low-performing nurses do not immediately reduce their error rate; instead, the improvement process requires gradual, long-term capability building.

Failed Tasks by the High-performing nurse: Since both Baseline and Training scenarios showed significant deviations from normality (Shapiro-Wilk p<0.001), the non-parametric Wilcoxon rank-sum test was applied. Results indicated no statistically significant difference in the number of failed tasks between the Baseline and Training scenarios (W=1675.5, p=0.5005). The mean number of failed tasks was similar across scenarios (Baseline: 1.92, Training: 2.08). These results suggest that the performance of the high-performing nurse remains stable between Baseline and Training, and the presence of trainers or the Training scenario itself does not significantly affect the number of failed tasks for this nurse.

Successful Tasks by the Low-performing nurse: Since the baseline group had identical values (all zeros), normality could not be assessed. The training group significantly deviated from normality (Shapiro-Wilk $W = 0.848, p = 2.72 \times 10^{-6}$), so the non-parametric Wilcoxon rank-sum test was applied. Results showed that the Training scenario achieved significantly more successful tasks ($W = 480, p = 9.86 \times 10^{-16}$) compared to the baseline. The mean difference was approximately 2.43 tasks per shift (Baseline: 0.00, Training: 2.43), indicating that training substantially improves the performance of low-performing nurses.

Successful Tasks by the High-performing nurse: Since both Baseline and Training scenarios exhibited significant deviations from normality (Shapiro-Wilk p<0.001 for both), the non-parametric Wilcoxon rank-sum test was applied. Results showed no statistically significant difference in the number of successful tasks between Baseline and Training (W=1958.5, p=0.403). The mean difference was minimal (Baseline: 16.67, Training: 16.72), indicating that training interventions to the High-performing nurse did not substantially affect task performance.

Nurse Utility by the Low-performing nurse: Since the Baseline group showed a significant deviation from normality (Shapiro-Wilk p=0.006) while the Training group did not (p=0.941), the non-parametric Wilcoxon rank-sum test was applied. Results showed that the Training scenario had significantly lower Nurse Utility ($W = 551.5, p < 2.72 \times 10^{-11}$) compared to the Baseline. The mean difference was approximately -10.07 units (Baseline: -9.97, Training: -20.03), indicating that the Training scenario substantially reduces the utility of the low-performing nurse.

Nurse Utility by the High-performing nurse: Since both groups showed deviations from normality (Shapiro-Wilk p<0.05 for Baseline and Training), the non-parametric Wilcoxon rank-sum test was applied. Results indicated no significant difference in utility between the Training and Baseline scenarios (W=1588, p=0.868). The mean utility values were 50.28 for Baseline and 47.95 for Training. This suggests that assigning a trainer nurse did not substantially change the utility of the high-performing nurse in the simulation.

Nurse Time Damage by the Low-performing Nurse: Since the Baseline group exhibited significant deviation from normality (Shapiro-Wilk, p=0.0021), the non-parametric Wilcoxon rank-sum test was applied. Results showed that the Training scenario did not reduce Nurse Time Damage (W=3347.5, p=1) compared to the Baseline. The mean time damage actually increased from 56.48 in Baseline to 140.33 s in Training. This outcome aligns with findings on failed tasks: although trainer nurses improve overall system throughput, low-performing nurses do not immediately reduce their error rate; instead, the improvement process requires gradual, long-term capability building. These results highlight that training interventions may not immediately mitigate the inefficiencies caused by underperforming staff.

Nurse Time Damage by the High-performing Nurse: Since both groups deviated significantly from normality (Shapiro-Wilk p<0.005), the non-parametric Wilcoxon rank-sum test was applied. Results showed no significant difference in Nurse Time Damage between the Training scenario and the Baseline (W= 1811.5,

p=0.5252). The mean Nurse Time Damage was 8.87 s in the Baseline and 8.92 s in the Training scenario, indicating that the good nurse's performance remained stable. This aligns with our previous observation that while the trainer nurses can improve overall system throughput, individual low-performing nurses require gradual capability building; high-performing nurses maintain consistent efficiency regardless of scenario.

Patient Distribution Across Doctors: In the Baseline scenario, some nurses show a preference for certain doctors, while in the Training scenario, the distribution is slightly more even. Using a Monte Carlo chi-square test for each simulation run, 18 out of 60 runs in the Baseline scenario showed a statistically significant nurse preference for specific doctors (p<0.05). In the Training scenario, 15 out of 60 runs showed a significant preference.

The variance in the number of patients served per doctor per run was 21.44 for the Baseline scenario and 26.29 for the Training scenario. A paired t-test comparing these variances found no statistically significant difference (p=0.838).

Conclusion: Nurses occasionally favor certain doctors in both scenarios, but the evidence is not strong enough to indicate a systematic preference. Patient distribution across doctors is generally uneven in both Baseline and Training scenarios, with no significant difference in variance between them.

**Conclusions**

The Training scenario substantially increases the number of patients served and reduces total patient delays demonstrating a clear improvement in system throughput under the use of CA Trust model. However, this comes at the cost of higher patient damage time and increased failed tasks by the low-performing nurse, highlighting that the error rate does not improve immediately. Instead, capability building for underperforming staff appears to be gradual and long-term in nature. The presence of trainer nurses stabilizes outcomes for high-performing nurses, whose performance and utility remain largely unaffected across scenarios.

From a managerial perspective, the Training scenario illustrates a critical trade-off: while system-level efficiency improves, short-term patient safety risks may increase due to persistent errors by low-performing nurses. Decision-makers should recognize that training interventions deliver operational gains but require sustained investment in long-term skill development to mitigate the risks associated with low-performing staff.

*Appendix B.4. Replacement (CA trust model) vs. Training (CA trust model)*

**Objective**

Compare two intervention strategies for handling low-performing nurses: immediate replacement vs skill development under training. Specifically, the study investigates how these interventions affect patient throughput, patient safety, nurse performance, and overall system efficiency.

**Design**

- Scenarios: Baseline vs. Training;
- Task selection policy: CA trust model applied in both scenarios;
- Metrics analyzed: Total patients served, total patient damage time, total patient delay, failed and successful tasks per nurse, nurse utility, nurse time damage, patient distribution across doctors.

**Statistical Approach**

All metrics were tested for normality using the Shapiro-Wilk test. Given that most metrics significantly deviated from normality, the non-parametric Wilcoxon rank-sum test was applied. For metrics with no variance in one group, the test was applied against the non-normal counterpart. Effect sizes, means, and p-values are reported. For patient distribution, Monte Carlo chi-square tests and variance comparisons via paired t-tests were conducted.

**Key Results**

Total Patients Served: Since both groups deviated significantly from normality (Shapiro-Wilk test, $p < 2.2 \times 10^{-16}$), the non-parametric Wilcoxon rank-sum test was applied. Results showed that the Replacement scenario served significantly more patients (W=104114, $p < 2.2 \times 10^{-16}$) compared to the Training scenario. The mean difference was approximately 12.47 patients per shift (Replacement: 41.58, Training: 29.12), indicating that immediate replacement of low-performing nurses results in higher patient throughput compared to training.

Total Patient Damage Time: Since both groups exhibited significant deviations from normality (Shapiro-Wilk test, Replacement: $W = 0.982, p = 0.00016$; $Training: W = 0.976, p = 6.73 \times 10^{-5}$ ), the non-parametric Wilcoxon rank-sum test was applied. Results showed that the Replacement scenario had significantly lower total patient damage time compared to the Training scenario ($W = 19695, p < 2.2 \times 10^{-16}$). The mean difference was approximately 53.3 seconds per shift (Replacement: 95.97 s, Training: 149.28 s), indicating that replacing low-performing nurses with more capable ones substantially reduces the overall damage exposure time of patients in the emergency department.

Total Patient Delay: Since both groups deviated significantly from normality (Shapiro-Wilk test, $p < 2.2 \times 10^{-16}$), the non-parametric Wilcoxon rank-sum test was applied. Results showed that the Replacement scenario had significantly lower patient delay (W=4440, $p < 2.2 \times 10^{-16}$) compared to the Training scenario. The mean difference was approximately 430.77 seconds per shift (Replacement: 6284.12, Training: 6714.88), indicating that immediate replacement of low-performing nurses reduces total patient delay more effectively than training.

Failed Tasks by the Low-performing nurse: Since the Replacement group deviated significantly from normality (Shapiro-Wilk test, $W = 0.908, p = 2.66 \times 10^{-4}$), while the Training group was approximately normal (W=0.967,p=0.102), the non-parametric Wilcoxon rank-sum test was applied. Results showed that the low-performing nurse had significantly fewer failed tasks in the Replacement scenario compared to the Training scenario (W=156.5, $p < 2.2 \times 10^{-16}$). The mean difference was approximately 3.88 failed tasks per shift (Replacement: 4.00, Training: 7.88), indicating that immediate replacement of low-performing nurses reduces task failures compared to training.

Failed Tasks by the High-performing nurse: Since both groups deviated significantly from normality (Shapiro-Wilk test, Replacement: W=0.910, p=0.0003; Training: W=0.860, $p = 6.3 \times 10^{-6}$), the non-parametric Wilcoxon rank-sum test was applied. Results showed no statistically significant difference between Replacement and Training (W=1727.5, p=0.695). The mean number of failed tasks was similar across conditions (Replacement: 1.92, Training: 2.08).

Successful Tasks by the Low-performing nurse: Since the Replacement group consisted of identical values (all zero) and the Training group deviated significantly from normality (Shapiro-Wilk test, $W = 0.848, p = 2.72 \times 10^{-6}$), the non-parametric Wilcoxon rank-sum test was applied. Results showed that the Training scenario produced significantly more successful tasks ($W = 480, p < 9.86 \times 10^{-16}$) compared to the Replacement scenario. The mean difference was approximately 2.43 successful tasks per shift (Replacement: 0.00, Training: 2.43), indicating that training substantially improved the performance of low-performing nurses compared to immediate replacement.

Successful Tasks by the High-performing nurse: Since both groups deviated significantly from normality (Shapiro-Wilk test, Replacement: $W = 0.813, p = 2.97 \times 10^{-7}$; Training: $W = 0.747, p = 7.98 \times 10^{-9}$), the non-parametric Wilcoxon rank-sum test was applied. Results showed that the Replacement scenario yielded significantly more successful tasks compared to the Training scenario (W=2269, p=0.013). The mean difference was approximately 1.20 tasks per shift (Replacement: 17.92, Training: 16.72). The results suggest that immediate replacement of the low-performing nurse leads to a modest but statistically significant improvement in the performance of the high-performing nurse, as measured by the number of successful tasks.

Nurse Utility by the Low-performing nurse: Since the Replacement group deviated significantly from normality (Shapiro-Wilk test, W=0.922, p=0.0009), while the Training group did not (W=0.991, p=0.94), the non-parametric Wilcoxon rank-sum test was applied. Results showed that the Replacement scenario yielded significantly higher utility for the low-performing nurse compared to the Training scenario ($W = 3038, p = 7.85 \times 10^{-11}$). The mean difference was approximately 9.57 utility points (Replacement: -10.47, Training: -20.03).

Nurse Utility by the High-performing nurse: Since both groups significantly deviated from normality (Shapiro-Wilk test, Replacement: W=0.951, p=0.018; Training: W=0.958, p=0.036), the non-parametric Wilcoxon rank-sum test was applied. Results showed no significant improvement of nurse utility in the Training scenario compared to Replacement (W=1189, p=0.999), with mean utility being higher under Replacement (56.45) than Training (47.95).

Nurse Time Damage by the Low-performing nurse: Since the Replacement group significantly deviated from normality (Shapiro-Wilk test, Replacement: W= 0.938, p=0.0045) while the Training group was approximately normal (W=0.974, p=0.230), the non-parametric Wilcoxon rank-sum test was applied. Results showed a significant increase in nurse time damage in the Training scenario compared to Replacement ($W = 3384, p < 2.2 \times 10^{-16}$), with mean time damage being substantially higher under Training (140.33) than Replacement (58.07).

Nurse Time Damage by the High-performing nurse: Since both groups significantly deviated from normality (Shapiro-Wilk test, Replacement: W=0.946, p=0.011; Training: W=0.937, p=0.004), the non-parametric Wilcoxon rank-sum test was applied. Results showed no significant difference in nurse time damage between the Training and Replacement scenarios (W=1750.5, p=0.604), with mean time damage being slightly lower under Training (8.92) than Replacement (9.22).

Patient Distribution Across Doctors: In the Replacement scenario, some nurses show a preference for certain doctors, while in the Training scenario, the distribution is slightly more even. Using a Monte Carlo chi-square test for each simulation run, 12 out of 60 runs in the Replacement scenario showed a statistically significant nurse preference for specific doctors (p<0.05). In the Training scenario, 15 out of 60 runs showed a significant preference.

The variance in the number of patients served per doctor per run was 29.71 for the Replacement scenario and 26.29 for the Training scenario. A paired t-test comparing these variances found no statistically significant difference (p=0.274).

Nurses occasionally favor certain doctors in both scenarios, but the evidence is not strong enough to indicate a systematic preference. Patient distribution across doctors is generally uneven in both Replacement and Training scenarios, with no significant difference in variance between them.

**Conclusions**

Replacement of low-performing nurses substantially improves patient throughput, reduces patient damage time, and lowers total patient delay compared to training. Training, however, enables low-performing nurses to achieve more successful tasks. The performance of high-performing nurses improves slightly under Replacement, while overall nurse utility favors Replacement. Patient distribution across doctors remains uneven but comparable across scenarios.

The Replacement scenario provides clear operational benefits, including faster patient service and reduced cumulative damage and delay. Training improves the skills of low-performing nurses, increasing their successful tasks but at the cost of higher patient delay and damage. Decision-makers must weigh the trade-off between immediate efficiency gains from Replacement versus long-term skill development through Training, especially when staffing resources are limited.